\documentclass[12pt,preprint]{aastex}

\begin{document}

\title{Ultraviolet Number Counts of Galaxies from \textit{Swift}
  UV/Optical Telescope Deep Imaging of the Chandra Deep Field
South} 

\author{E. A. Hoversten\altaffilmark{1}, C. Gronwall\altaffilmark{1}, D. E. Vanden Berk\altaffilmark{2},
  T. S. Koch\altaffilmark{1}, A. A. Breeveld\altaffilmark{3},
  P. A. Curran\altaffilmark{3}, D. A. Hinshaw\altaffilmark{4}, F. E. Marshall\altaffilmark{4},
  P. W. A. Roming\altaffilmark{1}, M. H. Siegel\altaffilmark{1}, and M. Still\altaffilmark{3}}

\altaffiltext{1}{Department of Astronomy \& Astrophysics, The Pennsylvania State
  University, 525 Davey Laboratory, University Park, PA 16801}
\altaffiltext{2}{Physics Department, St. Vincent College, Latrobe, PA 15650}
\altaffiltext{3}{Mullard Space Science Laboratory/UCL, Holmbury St. Mary, Dorking, Surrey RH5 6NT}
\altaffiltext{4}{NASA/Goddard Space Flight Center, Greenbelt, MD 20771}

\shorttitle{UV Number Counts of the CDF-S}

\begin{abstract}
Deep \textit{Swift} UV/Optical Telescope (UVOT) imaging of the
Chandra Deep Field South is used to measure galaxy number
counts in three near ultraviolet (NUV) filters (uvw2: 1928 \AA, uvm2:
2246 \AA, uvw1: 2600 \AA) and the $u$ band (3645 \AA).  UVOT
observations cover the break in the slope of the NUV number counts
with greater precision than the number counts by the Hubble Space
Telescope (HST) Space Telescope Imaging Spectrograph (STIS) and the
\textit{Galaxy Evolution Explorer} (\textit{GALEX}), spanning a range
from $21 \lesssim m_{AB} \lesssim 25$.  Number counts models confirm
earlier investigations in favoring models with an evolving galaxy
luminosity function.
\end{abstract}

\keywords{galaxies: evolution --- galaxies: UV properties ---
  galaxies: number density --- Chandra Deep Field South}

\section{Introduction}

Galaxy number counts as a function of magnitude provide direct
constraints on galaxy evolution in both luminosity and number density.
Number counts in the UV, in particular, can help trace the star
formation history of the universe.    Until recently obtaining faint
galaxy number counts in the UV has been difficult due to the small
areas surveyed
\markcite{Gardner00,Deharveng94,Iglesias04,Sasseen02,Teplitz06}({Gardner}, {Brown}, \&  {Ferguson} 2000; {Deharveng} {et~al.} 1994; {Iglesias-P{\'a}ramo} {et~al.} 2004; {Sasseen} {et~al.} 2002; {Teplitz} {et~al.} 2006).  While
the {\sl Galaxy Evolution Explorer (GALEX)} has allowed for the
measurement of UV galaxy number counts over a wide field of view
\markcite{Xu05}({Xu} {et~al.} 2005, $\sim 20$ deg$^2$), the confusion limit of {\sl GALEX}
restricts the magnitude range covered to 14 to 23.8 $m_{AB}$.  The
deepest UV number counts from {\sl HST\ } range from $m_{AB}$= 23 to
29 over an extremely small field of view of $\sim 1.3$ square arcminutes. 

Here, we present galaxy number counts obtained in 3 near UV filters
(1928 \AA, 2246 \AA, 2600\AA) as well as in the $u$ band (3645\AA)
obtained using the {\sl Swift\/} UV/Optical Telescope
\markcite{UVOT}(UVOT; {Roming} {et~al.} 2005).   Deep exposures were taken of a 289 square arcminute
field of view overlapping the Chandra Deep Field South
\markcite{CDFS}(CDF-S; {Giacconi} {et~al.} 2002) allowing for the measurement of number counts
from $m_{AB}$ = 21 to 26.  UVOT data covers the break in the slope of
the NUV number counts with greater precision than the existing  {\sl
  GALEX} and {\sl HST} number counts.  We use the UVOT number counts
to explore the evolution of star-forming galaxies out to $z\sim1$.

\section{Data \& Analysis \label{sect:data}}

The CDF-S was observed with UVOT, one of three telescopes onboard the
\textit{Swift} spacecraft \markcite{Swift}({Gehrels} {et~al.} 2004), the primary mission of which
is to study gamma-ray bursts (GRB).  The UVOT is a 30 cm telescope
with $f$-ratio 12.7 \markcite{UVOT}({Roming} {et~al.} 2005).  It has two grisms and seven
broadband filters.  The central wavelengths and widths of the uvw2,
uvm2, uvw1, and $u$ filters used in this paper can be found in Table
\ref{tab:uvotobs}.  For a  more detailed discussion of the filters, as
well as plots of the responses, see \markcite{UVOTcal}({Poole} {et~al.} 2008).

Observations of the CDF-S were made between July 7, 2007 and December
29, 2007.  CDF-S images are unbinned with a pixel scale of 0.5
arcseconds.  UVOT data processing is described in the UVOT Software
Guide\footnote{\texttt{http://heasarc.gsfc.nasa.gov/docs/swift/analysis}}.
The data were processed with a version of the UVOT pipeline in which
exposure maps are aspect corrected.  This feature is not currently
available for data currently in the archive but will appear in future
versions of the pipeline.  Image files and exposure maps were summed
using \texttt{UVOTIMSUM} from the publicly available UVOT FTOOLS
(HEAsoft
6.6.1)\footnote{\texttt{http://heasarc.gsfc.nasa.gov/docs/software/lheasoft/}}. 
This involves two flux conserving interpolations of the images, the
first to convert from the raw frame to sky coordinates, and the second
when summing the images.  Bad pixels are known and a correction is
applied.  UVOT, as is the case with all microchannel plate intensified
CCDs, is insensitive to cosmic rays.  The maximum exposure time in
each filter is given in Table \ref{tab:uvotobs}. 

Because \textit{Swift} is optimized for fast slewing to accomplish its GRB
mission, the pointing accuracy is of order 1 to 2 arcminutes.  In
addition the requirement that the solar panels face towards the Sun
causes the field of view to rotate over the course of the year.  As a
result the exposure times vary significantly across the summed images.
Exposure maps are nearly uniform in the center but become complicated
on the edges.  Table \ref{tab:uvotobs}  gives the area covered where
the exposure time is at least 98\% of the maximum exposure time in
each\ filter.  The 98\% value was chosen to
maximize the area used in this study while simultaneously maintaining
a magnitude limited sample. 

\begin{figure}
\plotone{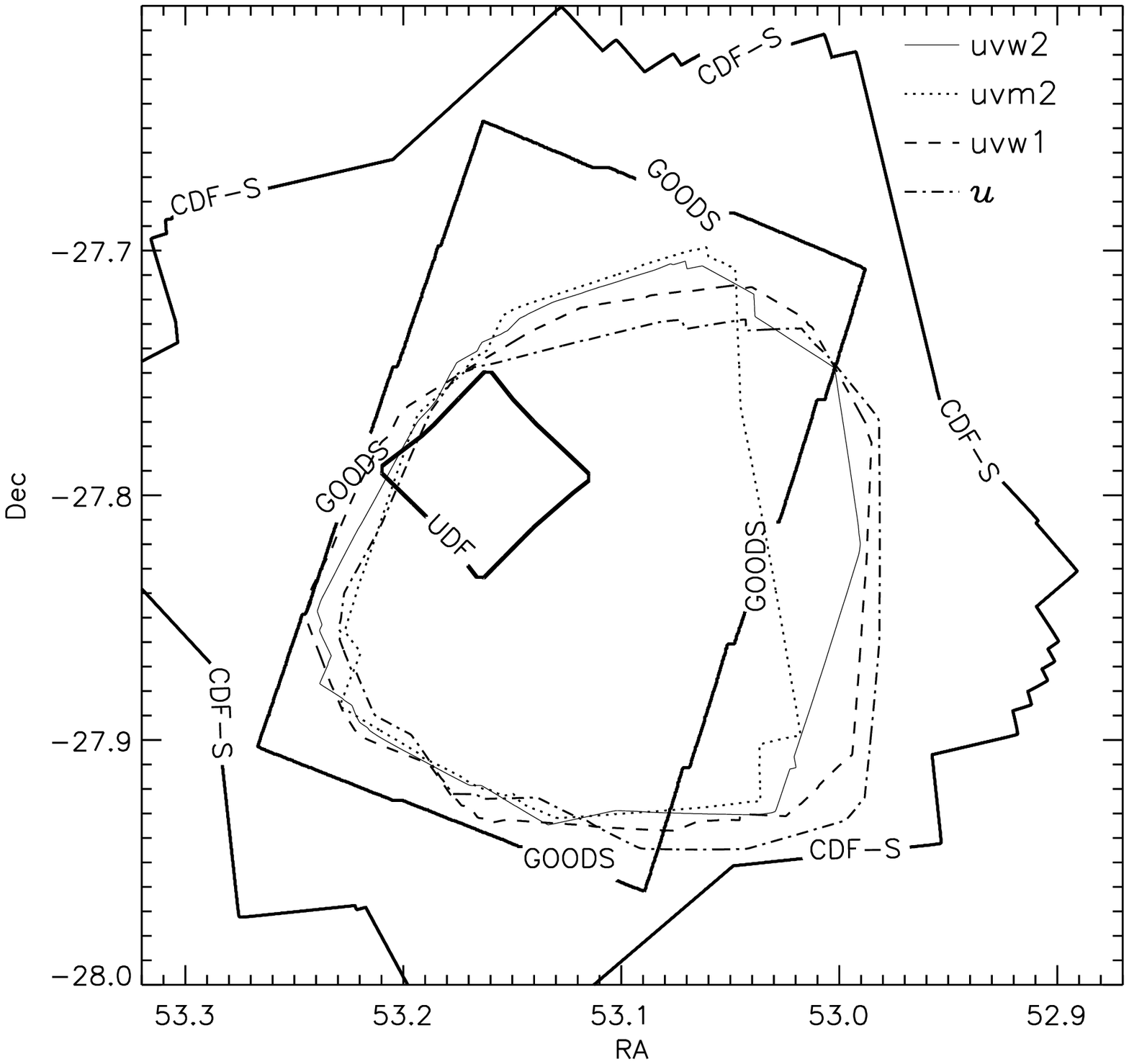}
\caption{Field of view for UVOT CDF-S observations.  Contours indicate
  the area covered with at least 98\% of the maximum exposure time as
  described in the text and tabulated in Table \ref{tab:uvotobs}.  The
  contours are uvw2 (thin solid line), uvm2 (dotted line), uvw1
  (dashed line), and $u$ (dot-dashed line).  For reference the extent
  of the Chandra Deep Field South \markcite{CDFS}({Giacconi} {et~al.} 2002), Hubble Ultra Deep
  Field \markcite{UDF}({Beckwith} {et~al.} 2006) and Great Observatories Origins Deep Survey
  \markcite{GOODS}({Giavalisco} {et~al.} 2004) are denoted by thick contours which are labeled.}
\label{fig:expmap}
\end{figure}

\begin{figure}
\plotone{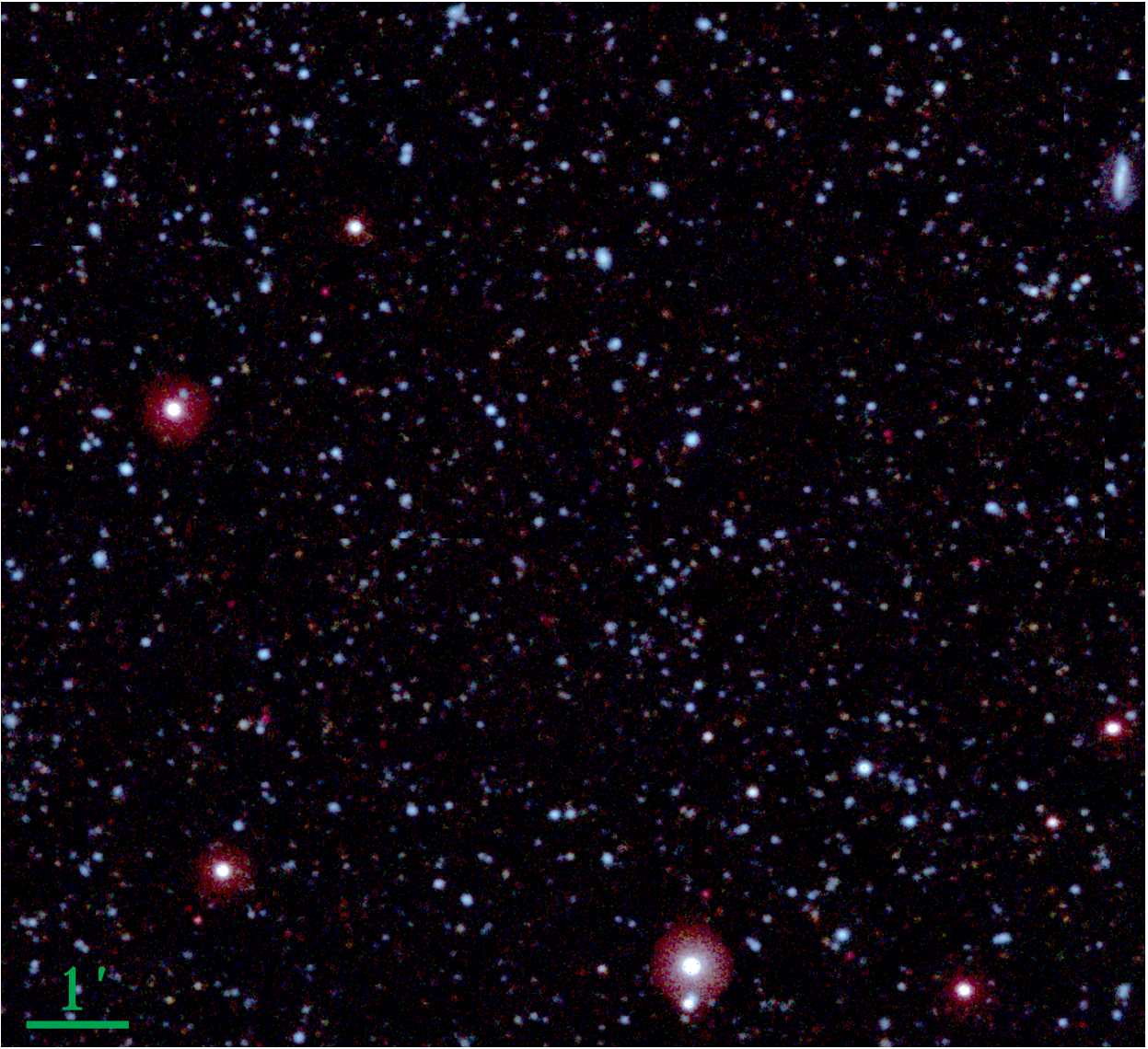}
\caption{Synthetic color image of a portion of the UVOT CDF-S deep
  field.  This image includes uvw2 (blue), uvm2 (green), and uvw1
  (red).  The $u$ band is not included in the image.  For reference
  the green bar is 1 arcminute long.}
\label{fig:ecdfs}
\end{figure}

The area in each filter covered by the 98\% exposure time criterion is
shown in Figure \ref{fig:expmap}.  For comparison the area covered by
the CDF-S, Hubble Ultra Deep Field \markcite{UDF}({Beckwith} {et~al.} 2006), and Great
Observatories Origins Deep Survey \markcite{GOODS}({Giavalisco} {et~al.} 2004) is shown by the
labeled contours.  A false color image of the central region of the
CDF-S using the uvw2, uvm2, and uvw1 images is shown in Figure
\ref{fig:ecdfs}. 

Photometry was performed using SExtractor \markcite{Sextractor}({Bertin} \& {Arnouts} 1996), a
publicly available code designed for the identification and
measurement of astrophysical sources in large scale galaxy survey
data.  A full listing of the SExtractor parameters used is provided in
the online version of Table \ref{tab:params}.  The background map,
which measures the local background due to the sky and other sources,
was generated internally by SExtractor.  To improve the detectability
of faint extended sources the filtering option was used with a
Gaussian filter.  The filter size was selected to match the full width
half maximum (FWHM) of the point spread function (PSF) as recommended
in the SExtractor manual.  The PSF was measured from the CDF-S image
for each filter using one star.  There was only one star which was
bright enough and isolated enough to accurately measure the outer
regions of the PSF.  The PSFs used were 3.30" in uvw2, 2.87" in uvm2,
2.86" in uvw1, and 2.67" in $u$.  Magnitudes were calculated from
\texttt{MAG\_AUTO} which is designed to be the best measure of the
total magnitudes of galaxies.  SExtractor was used to process count
rate images created by dividing the summed images by the exposure map.
The resulting output was converted to flux using the values given by
\markcite{UVOTcal}{Poole} {et~al.} (2008) for stellar spectra.  The fluxes were then converted
to AB magnitudes \markcite{Oke74}({Oke} 1974).  The number of sources detected in
each band ranges from 888 to 1260 and is given in Table
\ref{tab:uvotobs} along with the area covered in each image.

The UVOT detector is a microchannel plate intensified CCD which
operates in photon counting mode.  As such it is subject to
coincidence loss which occurs when two or more photons arrive at a
the same location on the detector within a single CCD readout interval
of 11 ms \markcite{Fordham00}({Fordham}, {Moorhead}, \&  {Galbraith} 2000).  When this happens only one photon will be
counted, which systematically undercounts the true number of photons.
The coincidence loss correction is at the 1\% level for m$_{AB} \sim
19$ in the UVOT filters we use in this paper.  For the magnitude ranges
considered in our number counts the coincidence loss is insignificant
and no attempts are made to correct for it.

By design the CDF-S is on a line of sight with very low Galactic
extinction.  In addition the area covered by the UVOT observation is
around 130 square arcminutes, depending on the filter, so variations
in extinction across the field are small.  According to the dust maps
of \markcite{Schlegel98}{Schlegel}, {Finkbeiner}, \&  {Davis} (1998) the range of Galactic extinction in our field is
$0.020 \leq {\rm A}_V \leq 0.030$.  Our photometry is corrected for
Galactic extinction based on the position of the source assuming the
Milky Way dust curve of \markcite{Pei92}{Pei} (1992).  The extinction correction is
largest in the uvm2 filter as it is centered on the 2175 \AA\ dust
feature which is pronounced in the Milky Way.  The extinction
correction ranges from $0.053 \leq {\rm A_{uvm2}} \leq 0.086$ across
the field in uvm2 which demonstrates that the extinction correction is
not a significant source of error in any of the filters. 

\section{Bias Corrections}

The raw number counts suffer from several biases which need to be
quantified.  Completeness addresses the inability to detect an object
either due to confusion with other sources  or limitations in the
photometry.  Eddington bias \markcite{Eddington13}({Eddington} 1913) occurs
because magnitude errors will preferentially scatter objects into
brighter magnitude bins because there are generally more objects at
fainter magnitudes.  There is also the potential for false detections of
objects due to noise.

These first two problems can be addressed simultaneously with a
Monte Carlo simulation, following the procedure set out in \markcite{Smail95}{Smail} {et~al.} (1995).
For each of the four images, synthetic galaxies were added and the
analysis repeated.  Synthetic galaxies were placed at random locations
on the image.  The magnitudes of the synthetic galaxies were between
21 and 27 in uvw2, uvm2 and uvw1 and between 20 and 25.5 in $u$ and
the relative numbers by magnitude follow the observed distribution
from the original SExtractor photometry in the relevant filter.  The
synthetic galaxies are given exponential profiles with semi-major axes
and ellipticities that match the observed distribution as a function
of magnitude.  Individual photon arrivals are modeled using Poisson
statistics and following the galaxy profile.  The resulting image is
then convolved with the UVOT PSF for the final image.

For each filter a single synthetic galaxy was added to the real image
and the photometry process described in \S \ref{sect:data} was
redone.  The resulting photometry catalog was checked to determine if
the synthetic galaxy was detected and at what magnitude.  This was
repeated 50,000 times for each filter to build up statistics on the
completeness.  The number counts were corrected by dividing by the
fraction of synthetic galaxies detected in the relevant magnitude bin.
These values are tabulated in Table \ref{tab:ncount}.  Following
\markcite{Smail95}{Smail} {et~al.} (1995) the number counts are truncated where the completeness
correction exceeds 50\%.  The Poisson error bars on the number counts
are also divided by the completeness correction to take into account
uncertainties introduced by the completeness correction.

Correcting for false detections was also done using the methods of
\markcite{Smail95}{Smail} {et~al.} (1995).  Using the exposure time and background count rate
calculated from the background map output by SExtractor noise frames
were simulated for each filter.  The photometry methods described in
\S \ref{sect:data} were repeated for each frame and the number of
false detections recorded as a function of magnitude.  For each filter
100 noise frames were analyzed.  The number of spurious sources,
$N_{spur}$, is shown as a function of magnitude for each filter in
Table \ref{tab:ncount}.  Out of all the simulated frames only one
spurious source was detected.  Given our deep exposures the
completeness correction truncates our number counts well before
background noise becomes an issue.

Galaxy number counts can be overestimated due to contamination by
Galactic stars and quasars.  The fraction of objects in the field that
are quasars is estimated by position matching the four UVOT photometry
catalogs with the Extended Chandra Deep Field South X-ray point source
catalog \markcite{ECDFS}({Lehmer} {et~al.} 2005).  Objects with X-ray detections are assumed to
be quasars.  The number of such sources in each band is 11, 11, 14,
and 21 for the uvw2, uvm2, uvw1, and $u$ bands respectively.  This
represents 1.2, 1.0, 1.1, and 2.3\% of the total sample.  These
sources have been removed from the number counts.  The number of AGN
per magnitude bin, $N_{AGN}$, is tabulated in Table \ref{tab:ncount}.

The problem of stellar contamination is greatly reduced by
the fact that the line of sight towards the CDF-S is out of the plane
of the Milky Way.  The CDF-S field was explicitly chosen to be
particularly sparse.  As a result the field is a statistical outlier,
and the stellar contamination in this field will be unusually low.  In
addition, the fraction of stars with significant UV flux is low,
particularly when the field points toward the Galactic halo where the
stellar population is very old.  This is another reason the stellar
contamination in the three NUV filters should be low.  

The contamination due to stars is estimated by position matching the
UVOT photometry catalogs with objects in the field with stellar
classifications in the COMBO-17 survey \markcite{Wolf04}({Wolf} {et~al.} 2004).  The
COMBO-17 survey includes photometry in 17 passbands for a $30\times30$
arcminute field surrounding the CDF-S.  It also contains photometric
redshifts and classifications of objects in the survey.  The UVOT
positions were compared with the objects classified as stars or white
dwarfs in COMBO-17.  This yields 24, 15, and 40 stars in the uvw2,
uvm2, and uvw1 NUV filters which corresponds to 2.7, 1.4, and 3.2\% of
the total sample. The number of stars per magnitude bin, $N_{star}$ is
shown in Table \ref{tab:ncount}.  Not all NUV number counts are
corrected for stellar contamination \markcite{Gardner00}(e.g. {Gardner} {et~al.} 2000).  Given
the numbers provided in Table \ref{tab:ncount} the number counts can
easily be recalculated without the stellar contamination correction.

However it is different in the $u$ band where more stars have
significant fluxes.  Position matching yielded 48 stars in the $u$
band which is 5.1\% of the total sample.  As in the NUV counts the
stellar contamination has been corrected for and the details are in
Table \ref{tab:ncount}.  \markcite{Capak04}{Capak} {et~al.} (2004) provide both raw number counts and
the number counts corrected for stellar contamination in the $U$ band
from observations around the Hubble Deep Field North (HDFN).  The $u$
and $U$ filters are comparable, and the HDFN is similar to the CDF-S
in being one of the darkest areas of the sky pointed out of the
Galactic disk with low Galactic extinction.   The level of stellar
contamination in \markcite{Capak04}{Capak} {et~al.} (2004) ranges from 66\% at $u=20$ to 6\% at
$u=25$.  At the bright end of this scale the values are comparable,
but at the faint end they are roughly twice as high as in the CDF-S.
One possible explanation for this discrepancy is that the
\markcite{Capak04}{Capak} {et~al.} (2004) sample covers $\sim 720$ arcsec$^2$ compared to 137
arcsec$^2$ for this sample.  Over this larger area one would expect
the number of stars to be closer to the average number expected for
that line of sight to the halo, while in our relatively smaller area
the number of stars can remain a statistical outlier.

Cosmic variance is another potential source of bias which arises due
to local inhomogeneities in the Universe.  Galaxies are known to
cluster on many different length scales.  As a result the density of
galaxies will differ along different lines of sight.  The smaller the
area covered by a survey the more the results will be biased by cosmic
variance.  A publicly available code from \markcite{Trenti08}{Trenti} \& {Stiavelli} (2008) was used to
estimate the errors due to cosmic variance in our number counts.  This
code is based in part on $N$-body simulations of galaxy structure
formation.  It uses the area of the survey, mean redshift, range of
redshifts observed and the number of objects detected to calculate the
error due to cosmic variance.  The mean redshift and redshift range of
each of our luminosity bins was estimated from the model number counts
described in \S \ref{sect:models}.  The results show that the
uncertainty due to cosmic variance are of the same order as the
Poisson errors for all of the filters and magnitude bins used here.
We therefore multiply our Poisson errors by a factor of $\sqrt 2$ to
take into account the effects of cosmic variance.

The resulting corrected number counts are shown in Figures
\ref{fig:ncuvw2} (uvw2), \ref{fig:ncuvm2} (uvm2), \ref{fig:ncuvw1}
(uvw1), and \ref{fig:ncuu} ($u$).  The number counts are also given in
Table \ref{tab:ncount}. In Figures \ref{fig:ncuvw2}, \ref{fig:ncuvm2},
and \ref{fig:ncuvw1} the number counts are plotted along side the NUV
number counts from GALEX \markcite{Xu05}({Xu} {et~al.} 2005) and STIS \markcite{Gardner00}({Gardner} {et~al.} 2000).  A
color conversion has been applied to shift the GALEX NUV filter and
STIS F25QTZ filter in the NUV channel by generating synthetic
magnitudes from a catalog of spectral synthesis models with a range of
ages and star formation histories and estimating the typical color
offset.  The GALEX and STIS NUV filters have very similar bandpasses
which typically differ by less than 0.01 magnitudes.  The uvm2 filter
has the tightest relationship with the NUV filters with a color
correction of 0.013.  The spread is larger in the uvw2 and uvw1
filters, but is still only of order 0.05.  In Figure \ref{fig:ncuu}
the UVOT $u$ band number counts are compared to the $U$ band counts of
\markcite{Capak04}{Capak} {et~al.} (2004) and \markcite{Eliche06}{Eliche-Moral} {et~al.} (2006) and the $u$ band measurements of
\markcite{Metcalfe01}{Metcalfe} {et~al.} (2001) and \markcite{Yasuda01}{Yasuda} {et~al.} (2001).  Color corrections to the UVOT
$u$ band were determined in the same fashion and are equal to 0.81 in
$U$ and 0.06 in $u$. 

\section{Models \label{sect:models}}

Simple models of number counts were constructed for both non-evolving
and evolving luminosity functions in the UVOT filters.  For each model
the luminosity function is summed over redshift.  This summation
includes two corrections.  The first is a filter correction to convert
the GALEX NUV filter to the UVOT uvw2, uvm2, and uvw1 filters and the
$U$ band to the UVOT $u$ filter.  The second is a $K$-correction to
convert the observed UVOT filter to the rest-frame UVOT filter.  Both
of these corrections are a function of redshift.

These corrections were calculated using a model galaxy spectrum
generated with the publicly available P\'{E}GASE spectral synthesis
code \markcite{PEGASE}({Fioc} \& {Rocca-Volmerange} 1997).  For the uvw2, uvm2, and uvw1 filters a starburst
galaxy model was used with a constant star formation rate, Solar
metallicity, and standard Salpeter IMF at an age of 800 Myr.  This was
chosen to match the model number counts in \markcite{Xu05}{Xu} {et~al.} (2005) which used the
SB4 starburst template of \markcite{Kinney96}{Kinney} {et~al.} (1996) because it most closely
matched the ratio of the local FUV to NUV luminosity densities
described by \markcite{Wyder05}{Wyder} {et~al.} (2005).  The P\'{E}GASE model is very nearly the
SB1 template of \markcite{Kinney96}{Kinney} {et~al.} (1996), however we model a range of internal
extinctions.  An $\Omega_M = 0.3$, $\Omega_\Lambda = 0.7$, $H_0=70$ km
s$^{-1}$ Mpc$^{-1}$ cosmology is used throughout. 

The $u$ band models were calculated assuming the cosmic spectrum of
\markcite{Baldry02}{Baldry} {et~al.} (2002) in addition to the starburst spectrum .  The cosmic
spectrum is a luminosity weighted average spectrum of galaxies with
$z\lesssim 0.1$ which makes it a good choice for a template
representative of all galaxies.  The empirical cosmic spectrum does
not extend far enough into the blue to be useful for modeling the UVOT
$u$ band let alone passing it through the filters at increasing
redshifts.  To extend the spectrum into the ultraviolet a template
spectrum was created in P\'{E}GASE from the best fitting parameters
given by \markcite{Baldry02}{Baldry} {et~al.} (2002).  The model number counts were corrected for
a range of models of internal extinction.  Models were calculated for
$0 \leq {\rm A}_V \leq 2$ for the Milky Way, LMC, and SMC dust models
of \markcite{Pei92}{Pei} (1992) and the starburst dust model of \markcite{Calzetti94}{Calzetti}, {Kinney}, \&  {Storchi-Bergmann} (1994).

Galaxy luminosity functions have traditionally been fit empirically
using Schechter functions \markcite{Schechter76}({Schechter} 1976).  The Schechter
function is given by
\begin{equation}
\phi(M){\rm d}M = \frac{\ln 10}{2.5} \phi^* \left [10^{0.4(M^* - M)} \right ]^{\alpha + 1} 
\exp  \left [-10^{0.4(M^* - M)} \right ] {\rm d}M
\end{equation}
where $\phi(M){\rm d}M$ is the number of galaxies with absolute
magnitude between $M$ and $M +{\rm d}M$ per Mpc$^3$.  Three free
parameters are fit using an empirical luminosity function; $\alpha$ is
the slope at the faint end of the luminosity function, $M^*$ is the
luminosity where the luminosity function turns over, and $\phi^*$ is
the density normalization.

For the non-evolving models the local galaxy luminosity function was
used at all redshifts.  The GALEX NUV galaxy luminosity function of
\markcite{Wyder05}{Wyder} {et~al.} (2005) was used for the uvw2, uvm2, and uvw1 models.  In the
$u$ band the models are based on the local $U$ band luminosity
function from \markcite{Ilbert05}{Ilbert} {et~al.} (2005).  In the evolving models the Schechter
function parameters $\alpha$, $\phi^*$, and $M^*$ vary with redshift.
In the uvw2, uvm2, and uvw1 bands the evolution of the Schechter
function parameters is based on their evolution at 1500 \AA\ as found
by \markcite{Arnouts05}{Arnouts} {et~al.} (2005), normalized to match the \markcite{Wyder05}{Wyder} {et~al.} (2005) NUV
parameters for the local universe.  For the $u$ band the evolution of
the Schechter function parameters comes from \markcite{Ilbert05}{Ilbert} {et~al.} (2005).  In
neither the non-evolving nor evolving models does the dust extinction
change as a function of redshift, nor does the underlying galaxy
template evolve.  A model with that level of complexity would be
beyond the scope of this paper.

The model number counts are also corrected for the Lyman forest and
continuum using the methods described by \markcite{Madau95}{Madau} (1995).  With the
exception of Hubble Deep Field $U$ band number counts
\markcite{Metcalfe01, Volonteri00}({Metcalfe} {et~al.} 2001; {Volonteri} {et~al.} 2000) NUV and $U$ model number counts
are generally not corrected for Lyman absorption.  Our modeling
reveals that this is justified.  In the bands considered in this paper this
affects the models by a few percent at 29th magnitude and
much less at brighter magnitudes.   Although the models described here
are plotted mainly for context the Lyman absorption corrections are
included. 

Example models are plotted with the number counts in Figures
\ref{fig:ncuvw2}, \ref{fig:ncuvm2}, \ref{fig:ncuvw1}, and \ref{fig:ncuu}.

\section{Results \label{sect:results}}

\begin{figure}
\plotone{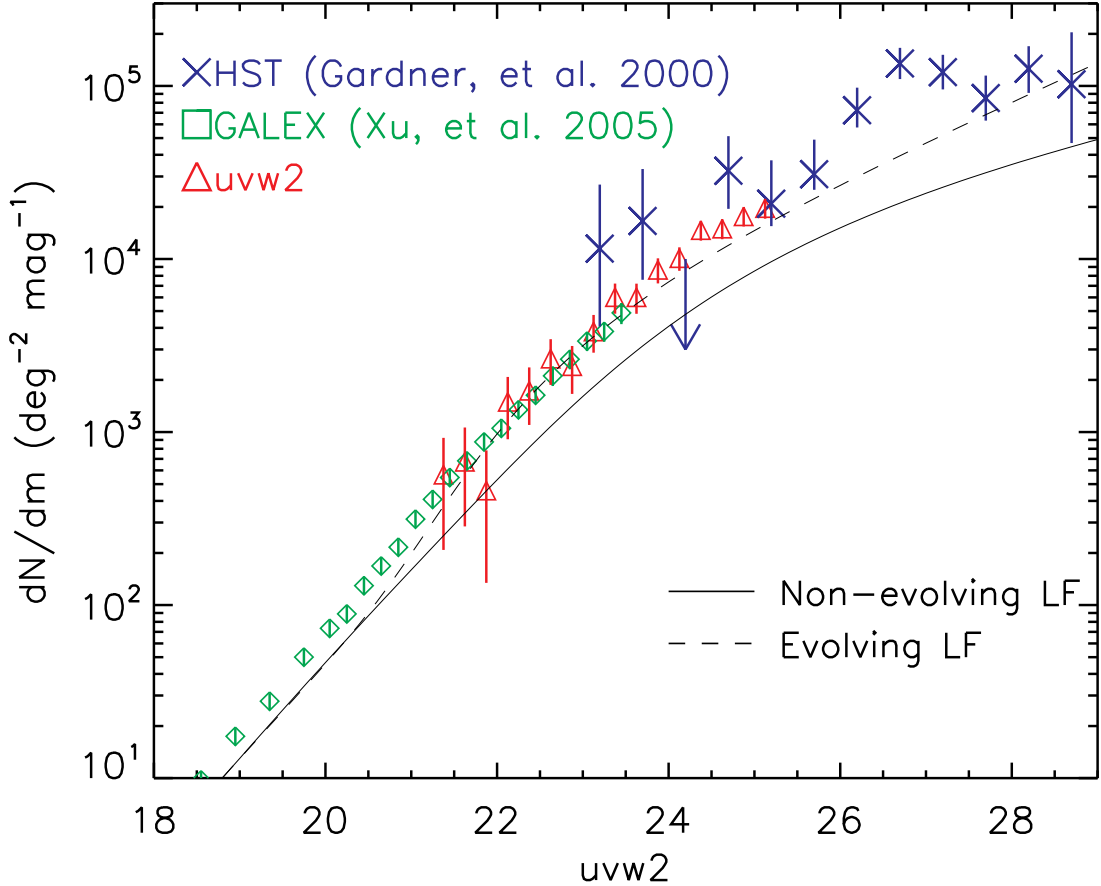}
\caption{UV number counts in the uvw2 filter (red triangles).  GALEX
  NUV number counts \markcite{Xu05}({Xu} {et~al.} 2005, green diamonds), and STIS NUV number
  counts \markcite{Gardner00}{Gardner} {et~al.} (2000, blue X's) are also plotted with a
  conversion to the uvw2 filter as described in the text.  Model
  number counts are also plotted for a starburst template galaxy and
  \markcite{Calzetti94}{Calzetti} {et~al.} (1994) dust model with $A_V=1$ and galaxy luminosity
  function parameters from \markcite{Wyder05}{Wyder} {et~al.} (2005).  Models for a non-evolving
  (solid line) and an evolving (dashed line) galaxy luminosity
  function following \markcite{Arnouts05}{Arnouts} {et~al.} (2005) are shown.}
\label{fig:ncuvw2}
\end{figure}

\begin{figure}
\plotone{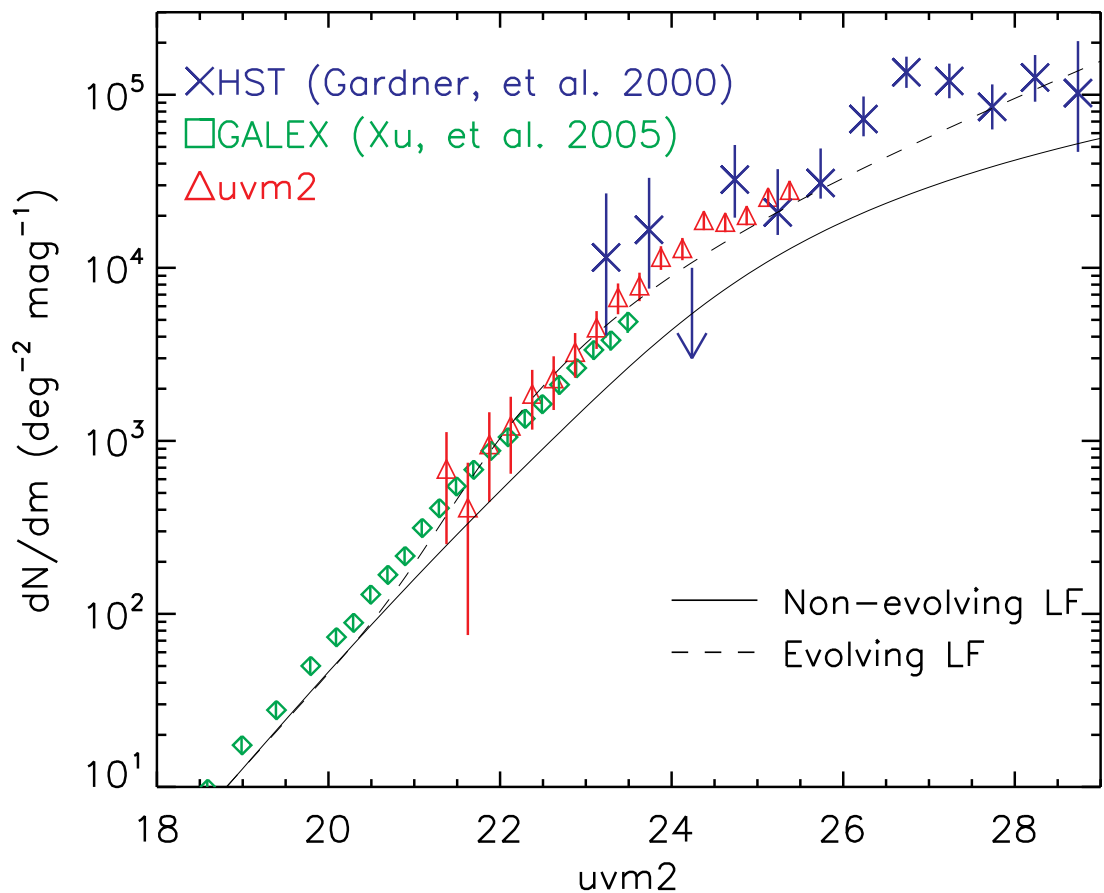}
\caption{UV number counts in the uvm2 filter (red triangles).  The
  rest of the description follows Figure \ref{fig:ncuvw2}.}
\label{fig:ncuvm2}
\end{figure}

Figures \ref{fig:ncuvw2} and \ref{fig:ncuvm2} show that in the uvw2
and uvm2 filters the number counts are in excellent agreement with the
NUV results from GALEX \markcite{Xu05}({Xu} {et~al.} 2005) and HST \markcite{Gardner00}({Gardner} {et~al.} 2000).
Furthermore, Figures \ref{fig:ncuvw2} and \ref{fig:ncuvm2} demonstrate
the unique contribution of UVOT.  The UVOT number counts have a
significant overlap with GALEX, however they continue $\sim 1.5$
magnitudes deeper with error bars comparable to those of GALEX.  In
this magnitude range they overlap with the HST number counts, but are
much less uncertain due to the wider field of view of UVOT as compared
to STIS.  While UVOT is not able to go as deep as HST, it provides
more precise number counts in the magnitude range where there is a
knee in the slope of the number counts. 

Figures \ref{fig:ncuvw2} and \ref{fig:ncuvm2} also show some of the
models discussed in \S \ref{sect:models}.  The models shown are for
the star forming galaxy template with \markcite{Calzetti94}{Calzetti} {et~al.} (1994) dust models
and ${\rm A_V} = 1.0$.  The solid line is a model with a non-evolving
galaxy luminosity function and the dashed line is an evolving model
following the evolution of the Schechter function parameters described
by \markcite{Arnouts05}{Arnouts} {et~al.} (2005).  The underlying models are the same in the two
figures, but have been calculated for the different filters.  In both
cases the non-evolving luminosity function model under-predicts the
number counts given the galaxy template and extinction assumptions.
However the evolving luminosity function model is simultaneously in
good agreement with the uvw2 and uvm2 number counts.  This is an
independent confirmation that the evolution in the luminosity function
parameters found by \markcite{Arnouts05}{Arnouts} {et~al.} (2005) are reasonable.

\begin{figure}
\plotone{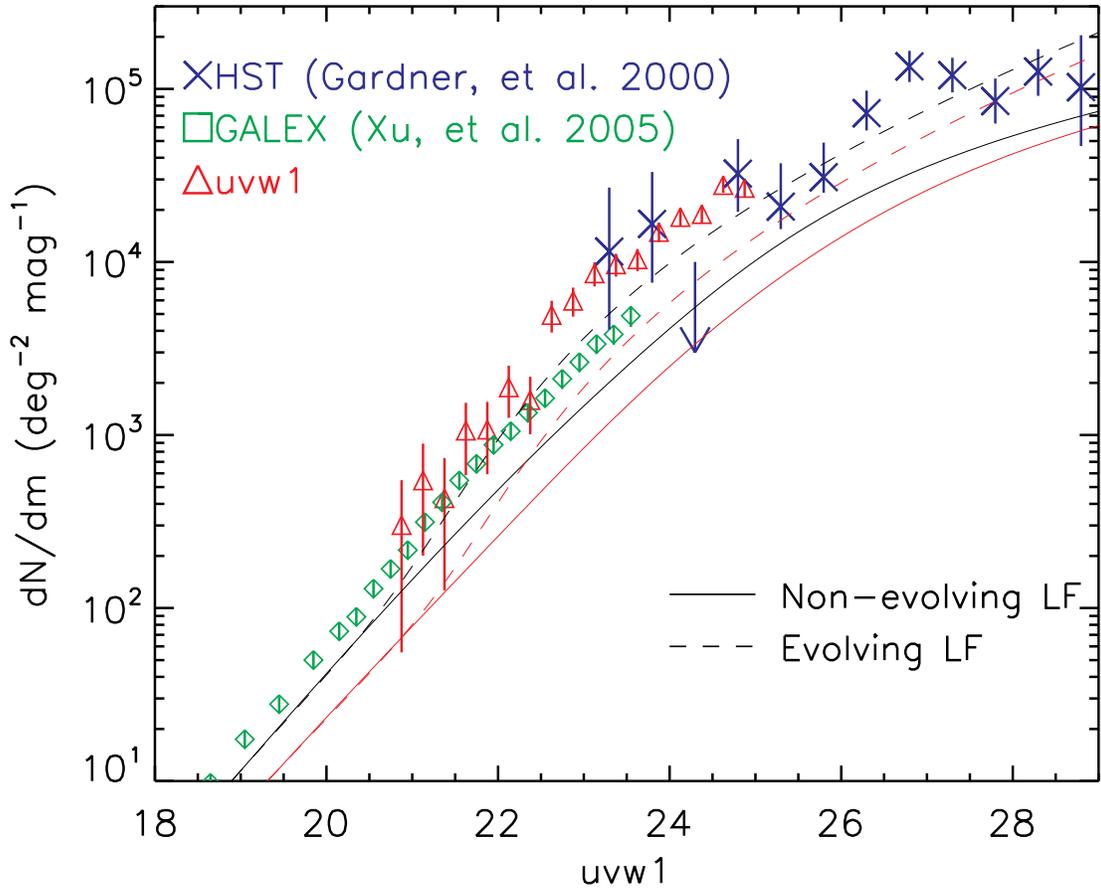}
\caption{UV number counts in the uvw1 filter (red triangles).  In
  addition to the description from Figure \ref{fig:ncuvw2}, model
  number counts assuming the cosmic spectrum of \markcite{Baldry02}{Baldry} {et~al.} (2002) as a
  template are shown in red for both non-evolving (solid line) and
  evolving (dashed line) luminosity functions.}
\label{fig:ncuvw1}
\end{figure}

Figure \ref{fig:ncuvw1} shows that the uvw1 number counts are
significantly higher than the GALEX NUV counts.  This can be explained
by the fact that the uvw1 filter has a tail in the red with
significant sensitivity between 3000 and 4000 \AA.  This extends
redward of the limits of the GALEX and STIS NUV filters.  At this
point bright elliptical galaxies can be detected in spite of the fact
that they do not produce an appreciable flux in the NUV.  Beyond the
extreme case of ellipticals, post-starburst galaxies with substantial
populations of A type stars and even to a lesser extent regular spiral
galaxies will also be over represented in the uvw1 number counts
compared to the NUV due to light being detected in the red wing of the
uvw1 filter.

The black models in Figure \ref{fig:ncuvw1} are the same as those in
Figures \ref{fig:ncuvw2} and \ref{fig:ncuvm2}.  The evolving
luminosity function model is still better than the no evolution model,
and is fairly representative of the GALEX and HST number counts.
However it does not agree with the uvw1 number counts as well as in
the uvw2 and uvm2.  This is due to the fact that the starburst galaxy
template is too blue to take into account the red objects which may be
detected by the red end of the uvw1 filter.  The red models assume the
same evolutionary parameters as the black models but uses the redder
cosmic spectrum of \markcite{Baldry02}{Baldry} {et~al.} (2002) as the galaxy template.  The models
using the cosmic spectrum template are below their respective star
forming template counterparts.  Thus the cosmic spectrum model has the
opposite problem in that it undercounts galaxies experiencing strong
star formation.  This shows that the simple modeling used here is less
successful for describing the uvw1 number counts, but also suggests
that the uvw1 filter could be useful in constraining the relative
numbers of different galaxy types over time.

\begin{figure}
\plotone{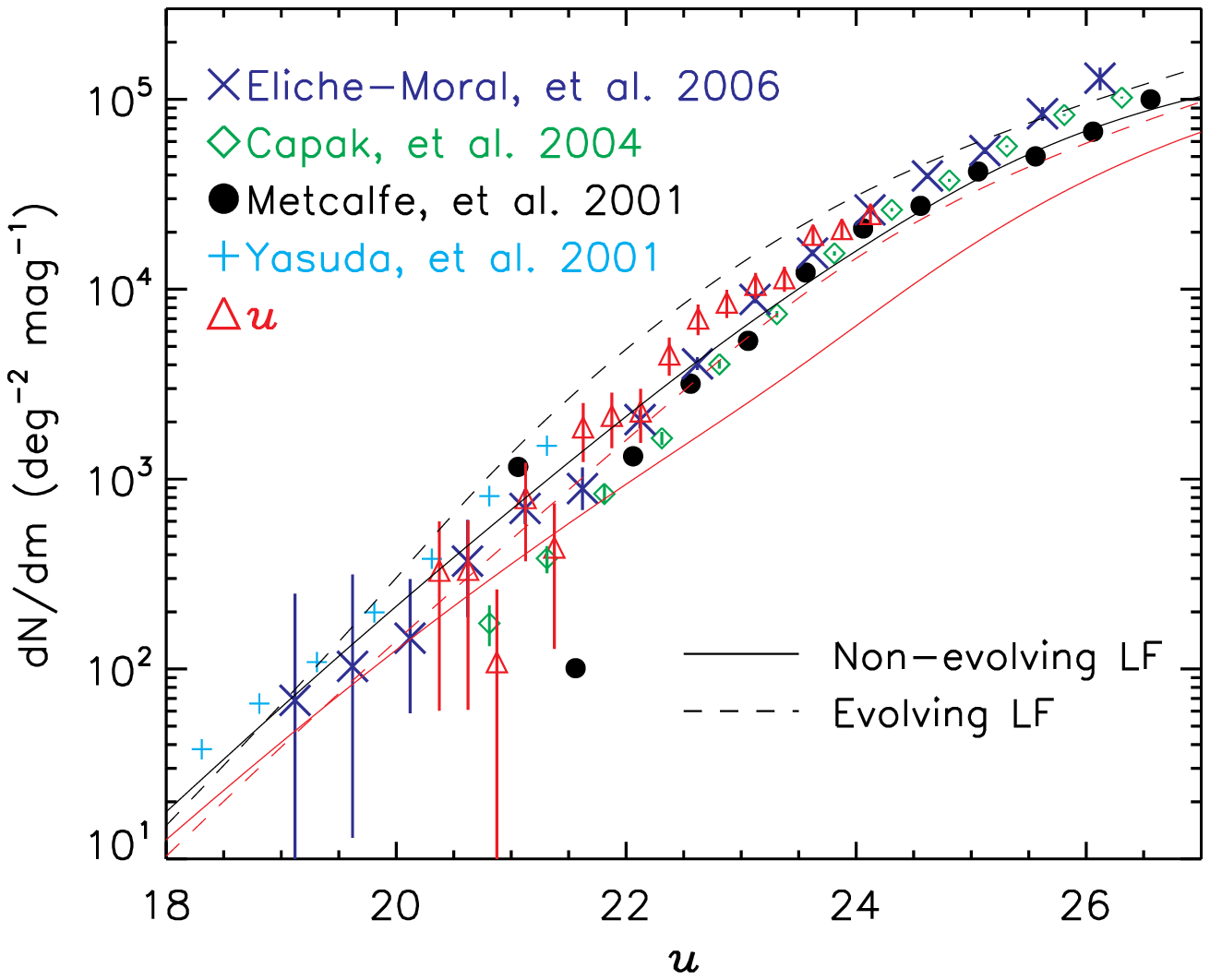}
\caption{Galaxy number counts in the UVOT $u$ filter (red triangles).
  $U$ number counts from \markcite{Capak04}{Capak} {et~al.} (2004) (green diamonds),
  \markcite{Eliche06}{Eliche-Moral} {et~al.} (2006) (blue X's), and \markcite{Metcalfe01}{Metcalfe} {et~al.} (2001) (black circles),
  as well as SDSS $u$ number counts from \markcite{Yasuda01}{Yasuda} {et~al.} (2001) (cyan plus
  signs) are also plotted with a conversion to the UVOT $u$ filter as
  described in the text.  \markcite{Yasuda01}{Yasuda} {et~al.} (2001) and \markcite{Metcalfe01}{Metcalfe} {et~al.} (2001) do not
  tabulate their errors.  Model number counts are also
  plotted for a starburst template galaxy and \markcite{Calzetti94}{Calzetti} {et~al.} (1994) dust
  model with $A_V=1$, and galaxy luminosity function parameters from
  \markcite{Ilbert05}{Ilbert} {et~al.} (2005) are shown in black for non-evolving (solid line) and
  an evolving (dashed line) galaxy luminosity functions.  Model number
  counts assuming the cosmic spectrum of \markcite{Baldry02}{Baldry} {et~al.} (2002) as a template
  are shown in red for both non-evolving (solid line) and evolving
  (dashed line) luminosity functions.}
\label{fig:ncuu}
\end{figure}

Figure \ref{fig:ncuu} shows that the $u$ band number counts are
generally in good agreement with other observations.  On the faint end
of the number counts the UVOT observations are in excellent agreement
with the $U$ band counts of \markcite{Capak04}{Capak} {et~al.} (2004) and \markcite{Eliche06}{Eliche-Moral} {et~al.} (2006) and the
$u$ band counts of \markcite{Metcalfe01}{Metcalfe} {et~al.} (2001).  At around magnitude 22 to 23
the UVOT number counts appear about 50\% higher.  One explanation for
this is that the \markcite{Yasuda01}{Yasuda} {et~al.} (2001) $u$ number counts are also higher
than the other observations on the faint end.  Modeling galaxy colors
shows that the SDSS $u$ is a much better proxy for UVOT $u$ than
Johnson $U$.  The higher number counts may be due to additional blue
sensitivity.  Figure \ref{fig:ncuu} also reveals that in the $u$
band UVOT does not have the unique advantage it has in the NUV filters
as it covers the same magnitude range as the ground based observations
and does not go as deep.  However it provides an independent check on
the ground based results. 

Figure \ref{fig:ncuu} also shows $u$ band model number counts for both
the starburst (black) and cosmic spectrum (red) templates, and both
non-evolving luminosity functions (solid) and those which evolve with
the parameters of \markcite{Ilbert05}{Ilbert} {et~al.} (2005).  In the $u$ band the evolving
luminosity function models with the starburst and cosmic spectrum
templates bracket the observed number counts, but then turn over at $u
\sim 25$ faster than the observed counts.

In summary, the UVOT is uniquely positioned to cover the knee in the
galaxy number counts compared to GALEX and HST in the NUV.  Due to its
smaller PSF it can go deeper than the GALEX confusion limit, and it's
larger field of view provides better statistics on the bright end of
the STIS number counts.  The simple model number counts used here
strongly point to an evolving galaxy luminosity function in agreement
with earlier studies.  More detailed models are needed to explain the
number counts in the uvw1 and $u$ filters, but are beyond the scope of
this paper.  However the measurements provided by this paper in the
magnitude range where the number counts turn over will enable a more
precise differentiation between models.  In addition, the three NUV
filters of UVOT are narrower than the single NUV filter of STIS and
GALEX so more color information is provided which is potentially
useful for more involved modeling.  Future plans include measurements
of the UV galaxy luminosity function as a function of redshift.

\acknowledgments

We acknowledge support from NASA Astrophysics Data Analysis grant,
\#NNX09AC87G.   This work is sponsored at PSU by NASA
contract NAS5-00136 and at MSSL by funding from the Science and
Technology Facilities Council (STFC).

{\it Facilities:} \facility{Swift (UVOT)}

\bibliography{}

\begin{deluxetable}{cccccc}
\tabletypesize{\scriptsize}
\tablecaption{\textit{Swift} UVOT observations of the CDF-S}
\tablewidth{0pt}
\tablehead{
\colhead{Filter} & \colhead{Central Wavelength (\AA)} & \colhead{FWHM (\AA)} &
\colhead{Exposure (s)} & \colhead{Area (arcmin$^2$)}\tablenotemark{a}& \colhead{$\#$ Sources}}
\startdata
uvw2 & 1928 & 657 & 144763 &  132.7 & 888\\
uvm2 & 2246 & 498 & 136286 &  112.0 & 1061\\
uvw1 & 2600 & 693 & 158334 &  143.2 & 1260\\
u    & 3465 & 785 & 124787 &  136.6 & 931\\
\enddata
\tablenotetext{a}{Area used for number counts where the exposure time is greater than or equal to 98\% of the maximum exposure time at the center of the image.}
\label{tab:uvotobs}
\end{deluxetable}

\begin{deluxetable}{ll}
\tabletypesize{\scriptsize}
\tablecaption{SExtractor parameters for CDF-S photometry}
\tablewidth{0pt}
\tablehead{\colhead{Parameter Name} & \colhead{Parameter Value}}
\startdata
\texttt{ANALYSIS\_THRESH} & 5.0\\
\enddata
\tablecomments{Table \ref{tab:params} appears in its entirety in the
  online version of the \textit{Astrophysical Journal}.  A portion is
  provided here for guidance regarding its form and content.}
\label{tab:params}
\end{deluxetable}

\begin{deluxetable}{ccccccccc}
\tabletypesize{\scriptsize}
\tablecaption{\textit{Swift} UVOT galaxy number counts in the CDF-S}
\tablewidth{0pt}
\tablehead{
\colhead{$m_{AB}$} & \colhead{Filter} & \colhead{Counts} & \colhead{$N_{raw}$} & \colhead{$N_{stars}$} & \colhead{$N_{AGN}$} & \colhead{$N_{spur}$} & \colhead{Completeness} & \colhead{Area}\\
 & & \colhead{(deg$^{-2}$ mag$^{-1}$)} & & & & & & \colhead{(arcmin$^2$)}}
\startdata
21.375 & uvw2 &   566 $\pm$   358 &    7 &  2 &  0 & 0.00 & 0.957 & 132.708 \\
21.625 & uvw2 &   673 $\pm$   388 &    6 &  0 &  0 & 0.00 & 0.967 & 132.708 \\
21.875 & uvw2 &   458 $\pm$   324 &    4 &  0 &  0 & 0.00 & 0.946 & 132.708 \\
22.125 & uvw2 &  1494 $\pm$   586 &   14 &  1 &  0 & 0.00 & 0.944 & 132.708 \\
22.375 & uvw2 &  1731 $\pm$   632 &   17 &  2 &  0 & 0.00 & 0.940 & 132.708 \\
22.625 & uvw2 &  2656 $\pm$   783 &   24 &  1 &  0 & 0.00 & 0.940 & 132.708 \\
22.875 & uvw2 &  2401 $\pm$   741 &   21 &  0 &  0 & 0.00 & 0.949 & 132.708 \\
23.125 & uvw2 &  3811 $\pm$   938 &   36 &  3 &  0 & 0.00 & 0.939 & 132.708 \\
23.375 & uvw2 &  6018 $\pm$  1191 &   56 &  2 &  3 & 0.00 & 0.919 & 132.708 \\
23.625 & uvw2 &  6011 $\pm$  1190 &   52 &  0 &  1 & 0.00 & 0.921 & 132.708 \\
23.875 & uvw2 &  8655 $\pm$  1432 &   76 &  2 &  1 & 0.00 & 0.915 & 132.708 \\
24.125 & uvw2 & 10108 $\pm$  1569 &   89 &  4 &  2 & 0.00 & 0.891 & 132.708 \\
24.375 & uvw2 & 14694 $\pm$  1921 &  117 &  0 &  0 & 0.00 & 0.864 & 132.708 \\
24.625 & uvw2 & 14980 $\pm$  1967 &  122 &  5 &  1 & 0.00 & 0.840 & 132.708 \\
24.875 & uvw2 & 17648 $\pm$  2223 &  129 &  1 &  2 & 0.00 & 0.775 & 132.708 \\
25.125 & uvw2 & 19683 $\pm$  2584 &  118 &  1 &  1 & 0.00 & 0.639 & 132.708 \\
\hline
21.375 & uvm2 &   687 $\pm$   435 &    5 &  0 &  0 & 0.00 & 0.943 & 110.982 \\
21.625 & uvm2 &   411 $\pm$   335 &    3 &  0 &  0 & 0.00 & 0.946 & 110.982 \\
21.875 & uvm2 &   953 $\pm$   509 &    7 &  0 &  0 & 0.00 & 0.952 & 110.982 \\
22.125 & uvm2 &  1221 $\pm$   575 &   12 &  3 &  0 & 0.00 & 0.956 & 110.982 \\
22.375 & uvm2 &  1866 $\pm$   705 &   14 &  0 &  0 & 0.00 & 0.973 & 110.982 \\
22.625 & uvm2 &  2293 $\pm$   786 &   17 &  0 &  0 & 0.00 & 0.962 & 110.982 \\
22.875 & uvm2 &  3259 $\pm$   940 &   26 &  2 &  0 & 0.00 & 0.955 & 110.982 \\
23.125 & uvm2 &  4510 $\pm$  1110 &   35 &  1 &  1 & 0.00 & 0.949 & 110.982 \\
23.375 & uvm2 &  6757 $\pm$  1365 &   49 &  0 &  0 & 0.00 & 0.941 & 110.982 \\
23.625 & uvm2 &  7894 $\pm$  1478 &   58 &  1 &  0 & 0.00 & 0.937 & 110.982 \\
23.875 & uvm2 & 11535 $\pm$  1801 &   82 &  0 &  0 & 0.00 & 0.922 & 110.982 \\
24.125 & uvm2 & 12975 $\pm$  1913 &   94 &  2 &  0 & 0.00 & 0.920 & 110.982 \\
24.375 & uvm2 & 18846 $\pm$  2337 &  130 &  0 &  0 & 0.00 & 0.895 & 110.982 \\
24.625 & uvm2 & 18397 $\pm$  2336 &  125 &  0 &  1 & 0.00 & 0.875 & 110.982 \\
24.875 & uvm2 & 20142 $\pm$  2537 &  131 &  2 &  3 & 0.00 & 0.812 & 110.982 \\
25.125 & uvm2 & 25642 $\pm$  3022 &  147 &  1 &  2 & 0.00 & 0.729 & 110.982 \\
25.375 & uvm2 & 28135 $\pm$  3647 &  126 &  3 &  4 & 0.00 & 0.549 & 110.982 \\
\hline
20.875 & uvw1 &   301 $\pm$   246 &    3 &  0 &  0 & 0.00 & 1.000 & 143.192 \\
21.125 & uvw1 &   545 $\pm$   345 &    6 &  1 &  0 & 0.00 & 0.921 & 143.192 \\
21.375 & uvw1 &   430 $\pm$   304 &    6 &  2 &  0 & 0.00 & 0.934 & 143.192 \\
21.625 & uvw1 &  1060 $\pm$   474 &   13 &  3 &  0 & 0.00 & 0.948 & 143.192 \\
21.875 & uvw1 &  1073 $\pm$   479 &   13 &  3 &  0 & 0.00 & 0.937 & 143.192 \\
22.125 & uvw1 &  1885 $\pm$   628 &   19 &  0 &  1 & 0.00 & 0.960 & 143.192 \\
22.375 & uvw1 &  1590 $\pm$   580 &   18 &  2 &  1 & 0.00 & 0.948 & 143.192 \\
22.625 & uvw1 &  4935 $\pm$  1029 &   50 &  4 &  0 & 0.00 & 0.937 & 143.192 \\
22.875 & uvw1 &  5968 $\pm$  1138 &   56 &  1 &  0 & 0.00 & 0.927 & 143.192 \\
23.125 & uvw1 &  8594 $\pm$  1376 &   81 &  2 &  1 & 0.00 & 0.913 & 143.192 \\
23.375 & uvw1 &  9689 $\pm$  1460 &   90 &  2 &  0 & 0.00 & 0.913 & 143.192 \\
23.625 & uvw1 & 10364 $\pm$  1519 &   98 &  3 &  2 & 0.00 & 0.902 & 143.192 \\
23.875 & uvw1 & 14905 $\pm$  1863 &  134 &  4 &  2 & 0.00 & 0.864 & 143.192 \\
24.125 & uvw1 & 18182 $\pm$  2092 &  154 &  3 &  0 & 0.00 & 0.835 & 143.192 \\
24.375 & uvw1 & 18908 $\pm$  2190 &  157 &  5 &  3 & 0.00 & 0.792 & 143.192 \\
24.625 & uvw1 & 27819 $\pm$  2775 &  206 &  3 &  2 & 0.00 & 0.727 & 143.192 \\
24.875 & uvw1 & 26781 $\pm$  3072 &  156 &  2 &  2 & 0.00 & 0.571 & 143.192 \\
\hline
20.375 &  $u$ &   328 $\pm$   268 &    5 &  2 &  0 & 0.00 & 0.962 & 136.569 \\
20.625 &  $u$ &   332 $\pm$   271 &    5 &  2 &  0 & 0.00 & 0.951 & 136.569 \\
20.875 &  $u$ &   108 $\pm$   153 &    4 &  3 &  0 & 0.00 & 0.972 & 136.569 \\
21.125 &  $u$ &   794 $\pm$   424 &    9 &  2 &  0 & 0.00 & 0.930 & 136.569 \\
21.375 &  $u$ &   436 $\pm$   308 &    4 &  0 &  0 & 0.00 & 0.967 & 136.569 \\
21.625 &  $u$ &  1871 $\pm$   642 &   22 &  5 &  0 & 0.00 & 0.958 & 136.569 \\
21.875 &  $u$ &  2154 $\pm$   699 &   21 &  2 &  0 & 0.00 & 0.930 & 136.569 \\
22.125 &  $u$ &  2269 $\pm$   717 &   24 &  3 &  1 & 0.00 & 0.929 & 136.569 \\
22.375 &  $u$ &  4532 $\pm$  1026 &   43 &  3 &  1 & 0.00 & 0.907 & 136.569 \\
22.625 &  $u$ &  7012 $\pm$  1280 &   62 &  0 &  2 & 0.00 & 0.902 & 136.569 \\
22.875 &  $u$ &  8493 $\pm$  1435 &   79 &  7 &  2 & 0.00 & 0.869 & 136.569 \\
23.125 &  $u$ & 10557 $\pm$  1628 &   88 &  3 &  1 & 0.00 & 0.839 & 136.569 \\
23.375 &  $u$ & 11420 $\pm$  1712 &   98 &  5 &  3 & 0.01 & 0.822 & 136.569 \\
23.625 &  $u$ & 19353 $\pm$  2280 &  152 &  4 &  4 & 0.00 & 0.785 & 136.569 \\
23.875 &  $u$ & 20726 $\pm$  2409 &  155 &  6 &  1 & 0.00 & 0.753 & 136.569 \\
24.125 &  $u$ & 24928 $\pm$  2850 &  160 &  1 &  6 & 0.00 & 0.647 & 136.569 \\
\enddata
\label{tab:ncount}
\end{deluxetable}

\end{document}